\documentclass[journal=jacsat,manuscript=article]{achemso}
\setkeys{acs}{usetitle = true}

\usepackage{amssymb}
\usepackage[version=3]{mhchem} 
\author{Jennifer M. Rieser}
\email{jrieser@sas.upenn.edu}
\affiliation[University of Pennsylvania]
{Department of Physics and Astronomy, University of Pennsylvania, Philadelphia, PA 19104-6396, USA}

\author{P. E. Arratia}
\affiliation[University of Pennsylvania]
{Department of Mechanical Engineering and Applied Mechanics, University of Pennsylvania, Philadelphia, PA 19104-6396, USA}

\author{A. G. Yodh}
\affiliation[University of Pennsylvania]
{Department of Physics and Astronomy, University of Pennsylvania, Philadelphia, PA 19104-6396, USA}

\author{J. P. Gollub}
\affiliation[Haverford College]
{Department of Physics, Haverford College, 370 Lancaster Avenue, Haverford, PA 19041, USA}
\alsoaffiliation[University of Pennsylvania]
{Department of Physics and Astronomy, University of Pennsylvania, Philadelphia, PA 19104-6396, USA}

\author{D. J. Durian}
\email{djdurian@physics.upenn.edu}
\affiliation[University of Pennsylvania]
{Department of Physics and Astronomy, University of Pennsylvania, Philadelphia, PA 19104-6396, USA}

\title{Tunable capillary-induced attraction between vertical cylinders }
\abbreviations{IR,NMR,UV}
\bibstyle{achemso}

\begin{document}

\begin{tocentry}
 \centering
  \includegraphics[width = 7.25cm]{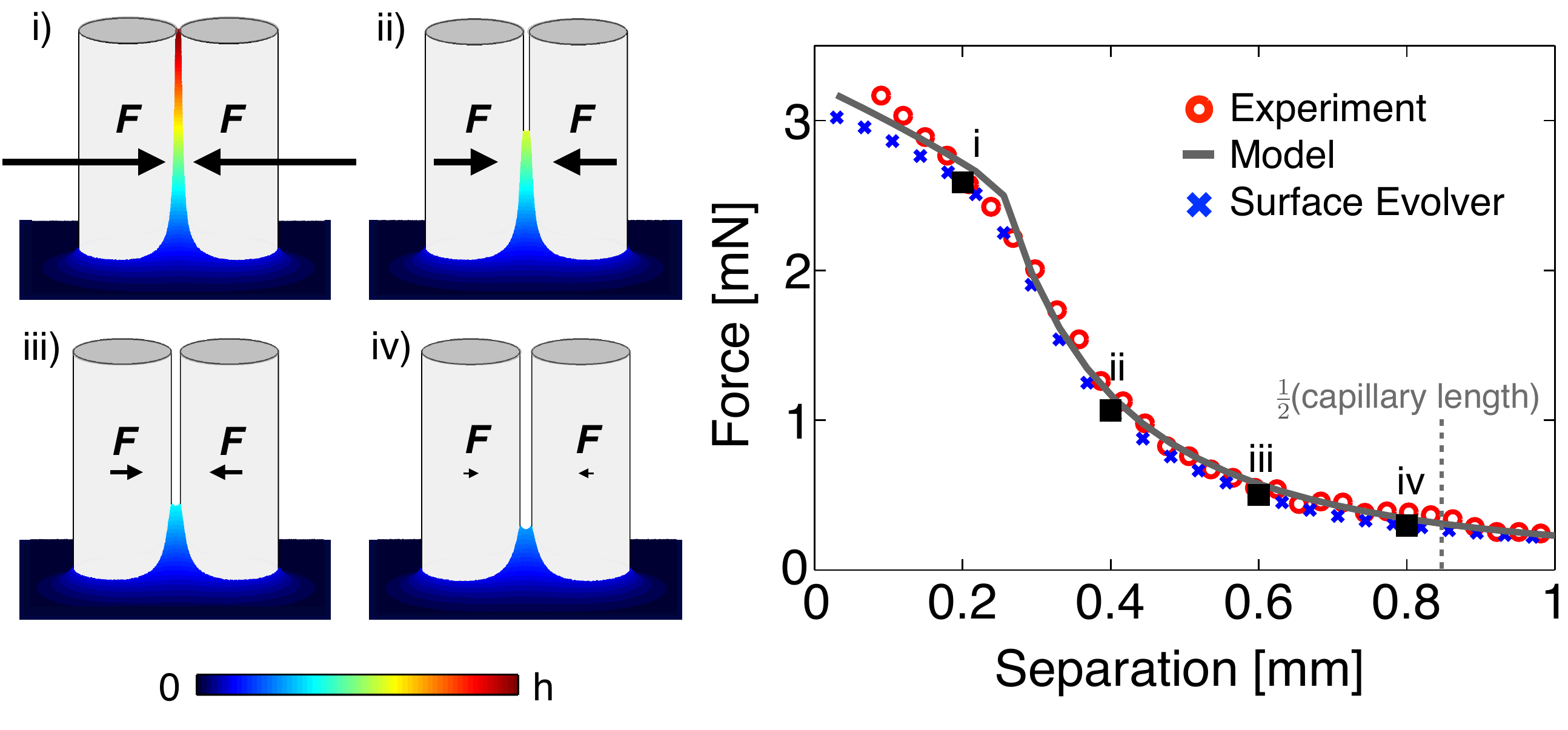} 
\end{tocentry}

\begin{abstract}
Deformation of a fluid interface caused by the presence of objects at the interface can lead to large lateral forces between the objects.  We explore these fluid-mediated attractive force between partially submerged vertical cylinders.   Forces are experimentally measured by slowly separating cylinder pairs and cylinder triplets after capillary rise is initially established for cylinders in contact.  For cylinder pairs, numerical computations and a theoretical model are found to be in good agreement with measurements.  The model provides insight into the relative importance of the contributions to the total force.  For small separations, the pressure term dominates, while at large separations, surface tension becomes more important. A cross-over between the two regimes occurs at a separation of around half of a capillary length.  The experimentally measured forces between cylinder triplets are also in good agreement with numerical computations, and we show that pair-wise contributions account for nearly all of the attractive force between triplets.  For cylinders with equilibrium capillary rise height greater than the height of the cylinder, we find that the attractive force depends on the height of the cylinders above the submersion level, which provides a means to create precisely-controlled tunable cohesive forces between objects deforming a fluid interface.
\end{abstract}

\section{Introduction}
Flow properties of granular materials can be greatly influenced by the presence of a small amount of fluid\cite{Herminghaus:2005ei,Mitarai:2006dj}.  This fluid-driven change in behavior can be quite dramatic and has important implications for industrial processing, mining, and construction, as well as geological phenomena such as landslides.  Nevertheless, an understanding of how local capillary-bridge-induced force distributions influence bulk flow properties and give rise to global deformation is lacking.  

Previous studies have explored how global mechanical stability and flow response vary with liquid content\cite{Hornbaker:1997ws,Halsey:343475,Bocquet:1998wt,Mason:1999tr,Samadani:2001eq,Tegzes:2002dp,Nowak:2005gf}, but relating global response to microscopic details has proven challenging in three-dimensional systems.  While X-ray tomography provides detailed information about the three-dimensional structure of the distribution of liquid inside the granular material\cite{Scheel:2008hh}, little progress has been made in three-dimensional systems towards controlling where liquid resides throughout the granular material, making systematic exploration of the relationship between grain-scale structure and large-scale flows challenging.

In two-dimensional rafts of floating particles, however, the fluid distribution is uniform, and fluid-mediated interactions have been characterized for a variety of particles\cite{Nicolson:1949up,Bowden:1997js,Mansfield:1997wg,Vassileva:2005ed,Vella:2005hi,Loudet:2005be,Vella:2006bq,Madivala:2009gl,Lewandowski:2010hw,Berhanu:2010gm,Schwenke:2014jm,Xue:2014bw}.  Additionally, it has recently been shown that the fluid distribution in a particle monolayer in a water-lutidine mixture can be controlled and uniform.\cite{Gogelein:2010fk}   In this paper, we characterize the capillarity-induced interactions between vertical cylinders standing upright on a substrate in a pool of liquid.  Here the fluid is distributed uniformly, as in rafts and monolayers.  Furthermore, the strength of the attractive force can be tuned by varying the depth of the pool of fluid.  Both the uniformity and tunability of the these forces in two dimensional systems may prove helpful for understanding the influence of local fluid-grain interactions on bulk-scale granular flow. 

Wetting and capillary interactions have long been studied.\cite{QuereBook,Pomeau:2006cs,Bonn:2009ha}  For vertical cylinders, one context is surface roughness and superhydrophobicity due to an array of micropillars \cite{Quere:2008cw,Herminghaus:2000,Lau:2003hj,Bernardino:2012gt}.  If the micropillars are long and flexible, elastocapillary effects can lead to coalescence, which has important implications in nature as well as engineering and materials science. \cite{Chandra:2009gh,Pokroy:2009uj,Chandra:2010eo,Bernardino:2010hq,Kang:2011jf,Taroni:2012gy}  In this paper, however, we will focus on a different context.  Here the cylinders are rigid and are not anchored to the substrate on which they sit. As a result, the cylinders do not bend or deform but are free to move laterally in response to the fluid forces.

In this latter context, Princen\cite{Princen:1969ty} considered the wicking behavior of long thin fibers and developed a model to estimate the capillary rise height of liquid between two rigid vertical cylinders as a function of their separation.  Kralchevsky, \textit{et al.}\ \cite{Kralchevsky:1992vn,Kralchevsky:1993wc,KralchevskyBook} solved the linearized Laplace equation to derive an analytical form for the lateral forces between floating colloidal particles in the limit of small deformations of the fluid surface.  Velev, \textit{et al.}\ \cite{Velev:1993wz} and Dushkin, \textit{et al.}\ \cite{Dushkin:1996hu} used a torsion balance to experimentally measure the lateral forces between two partially submerged sub-millimeter-diameter vertical cylinders at separations greater than half a capillary length.  Forces at these separations were shown to be similar to the predictions of Kralchevsky, \textit{et al.}\ \cite{Kralchevsky:1992vn,Kralchevsky:1993wc,KralchevskyBook}, indicating that the small-deformation approximation is reasonable at large separations.  Cooray, \textit{et al.}\ \cite{Cooray:2012bn} later achieved even better agreement with the experimental values by numerically solving the full non-linear Laplace equation. However, all previous works\cite{Kralchevsky:1992vn,Kralchevsky:1993wc,KralchevskyBook,Cooray:2012bn,Velev:1993wz,Dushkin:1996hu} characterizing these forces have been restricted to sub-millimeter-diameter cylinders of effectively infinite height as cylinder height always exceeds the equilibrium capillary rise height.  Further, previous experimental works\cite{Velev:1993wz,Dushkin:1996hu} only characterize forces for separations larger than half a capillary length.  The capillary attraction of vertical cylinders of finite height in the millimeter-diameter range has not been investigated experimentally or theoretically.

In this paper, we explore the fluid-mediated attractions between rigid vertical cylinders of finite height and diameter larger than the capillary length.  A custom-built apparatus permits measurement of forces between several pairs of vertical cylinders as they are quasi-statically separated.  We thus measure forces for separations as small as $80$~$\mu$m.  The fluid-surface deformations are large at these small separations, and therefore the analytical form obtained from the linearized Laplace equation \cite{Kralchevsky:1992vn,Kralchevsky:1993wc,KralchevskyBook} is not valid.  However, we find reasonable agreement with an extension of the Princen model to calculate the lateral forces between vertical cylinders.  Numerical computations are also shown to be in good agreement with experimental measurements.  We observe that for cylinders of finite height at small separations, the capillary rise of the fluid reaches the tops of the cylinders, thereby introducing a way to control the strength of cohesion between cylinders.  Lastly, we observe a velocity-dependent hysteresis consistent with the observations of Velev, \textit{et al.}\,\cite{Velev:1993wz}.

\section{Methods}
\subsection{Experimental Setup}
\begin{figure}[h!]
 \centering
  \includegraphics[width=5in]{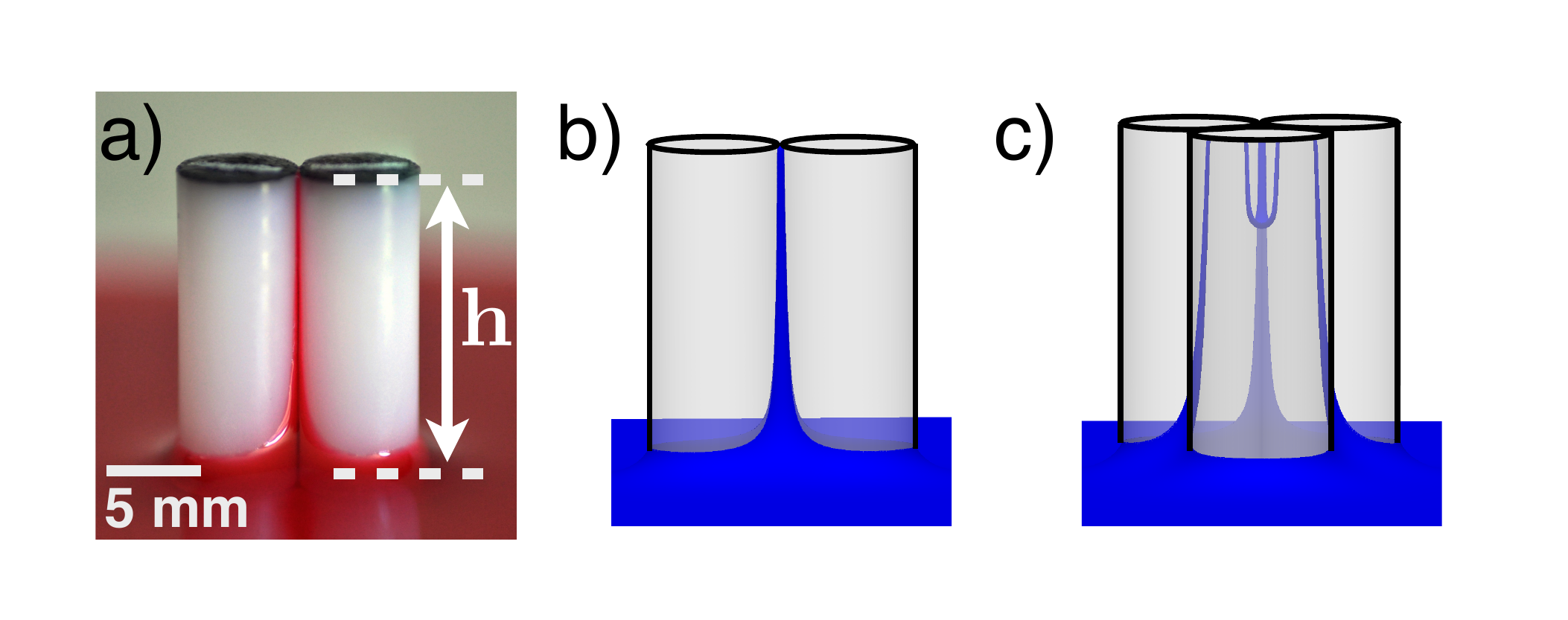} 
  \caption{a) A pair of $R=3.175$~mm upright cylinders standing in mineral oil (dyed red) viewed from the side.  b) Final Surface Evolver output for a pair of cylinders with similar conditions to a), also viewed from the side.  c)  Final Surface Evolver state for a group of three cylinders.    The capillary rise reaches the tops of the cylinders at small separations, causing the resulting cohesive force between to be set by $h$, the exposed height of the cylinders above the liquid reservoir. }
  \label{fig:capRise}
\end{figure}

We measure the capillarity-induced attractive forces between pairs and triplets of vertical cylinders partially submerged in a fluid, as shown in Figure~\ref{fig:capRise}a.  The cylinders are acetal dowel pins with density $\rho_{cyl} = 1410$~kg$/$m$^3$, height $H =19.05$~mm,  and radius $R = 3.175$~mm. The fluid is heavy viscosity mineral oil with density $\rho = 870 \pm 10$~kg/m$^3$.  The acetal-air-mineral oil contact angle, $\theta_c$, is estimated to be $\theta_c = 20 \pm 5^{\circ}$ from numerous photographs of a single cylinder partially submerged in oil.  Using the equation for capillary rise inside a cylindrical tube, $h_{\mathrm{rise}} = 2 \gamma \cos \theta_c/ (\rho g r_{\mathrm{tube}})$,  the surface tension is estimated to be $\gamma = 27.1 \pm 0.5$~dyn/cm from photographic measurements of capillary rise heights inside capillary tubes of both $5$ $\mu$L and $50$ $\mu$L volumes.  Most of the uncertainty in the surface tension measurement results from the uncertainty in the contact angle.  The capillary length of the oil is $l_c =(\rho g/\gamma)^{-1/2}= 1.8 \pm 0.2$~mm.     

\begin{figure}[h!]
  \centering
  \includegraphics[width=5in]{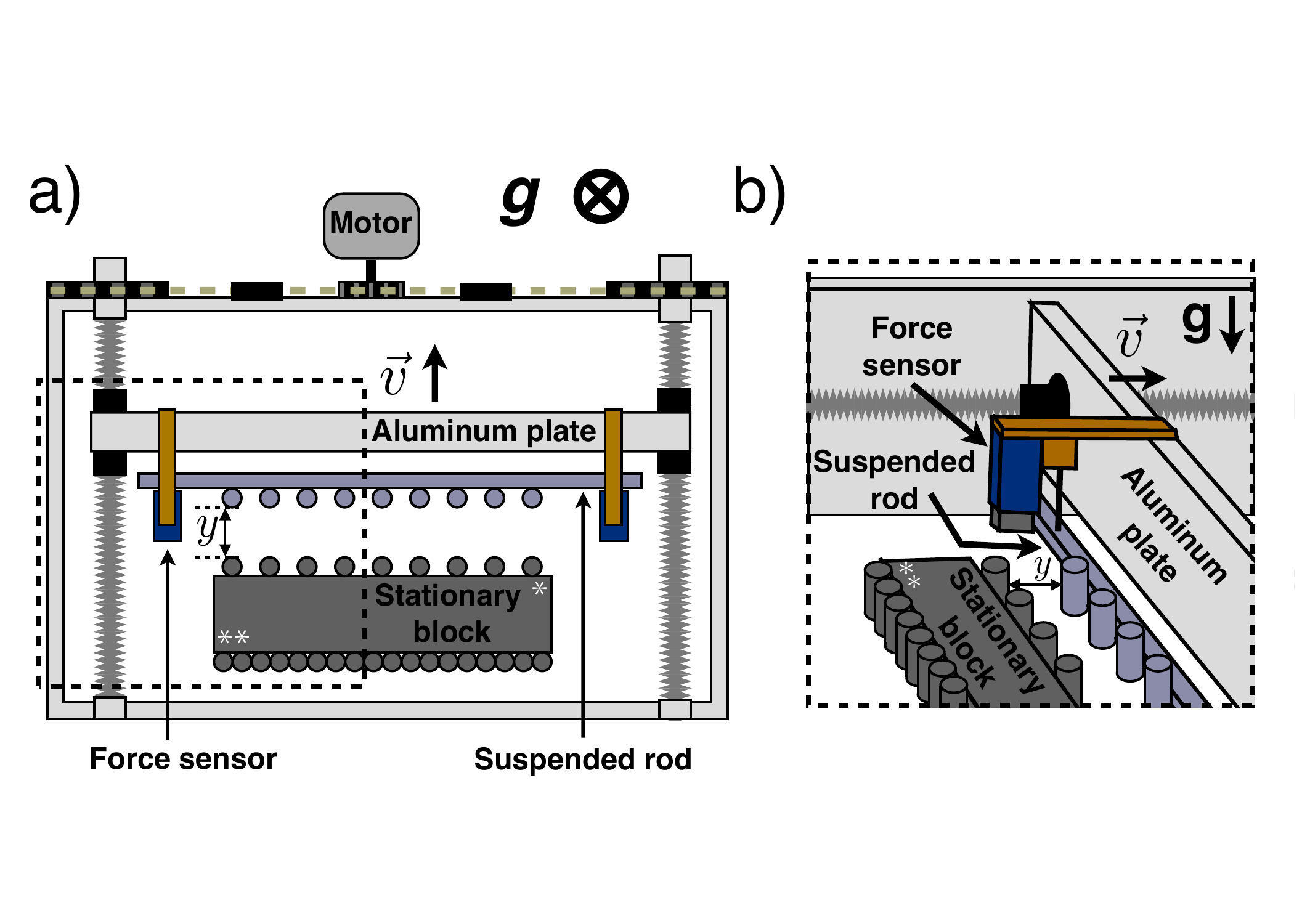} 
  \caption{a) Top-down view of the setup for measuring cohesive forces between cylinder pairs (single white star in the upper right corner of the stationary block, as shown) and cylinder triplets (the stationary block is rotated $180^{\circ}$ so that the two white stars are located in the upper right corner).  This schematic not to scale: in the experiments, each cylinder has radius $R = 3.175$~mm and  there are $15$ cylinder pairs and $16$ cylinder triplets.  Another row of cylinders is glued to a rod suspended from a plate attached to a motor.  A known amount of mineral oil is added to the surrounding container for each set of experiments. Initially, the cylinders attached to the suspended rod are moved into contact with the cylinders attached to the stationary block, allowing capillary bridges to form between them.  The suspended cylinders are then pulled away from the stationary cylinders.  Lateral capillary forces resist this motion, causing the suspended rod to come into contact with and exert a force on the force sensors. b) Side view of the region enclosed by the dashed box in a). }
  \label{fig:cohesionSchematic}
\end{figure}

A custom-built apparatus, shown in Figure~\ref{fig:cohesionSchematic}, is employed to measure oil-induced cohesive interactions between pairs and  triplets of identical upright cylinders.  Two threaded rods are mounted to the surrounding liquid-tight box with only the freedom to rotate. This rotation is driven by a stepper motor at a constant rate, permitting the translational motion of aluminum plate held by the threaded rods.  Two force sensors mounted to the aluminum plate, one at each end, are sensitive to deflections perpendicular to the long axis of the plate.  Equally spaced vertical cylinders are glued to a rod suspended from the aluminum plate.  Neighboring cylinders attached to this rod have center-to-center separations of $4R$, and all cylinder bases are about $1$~mm above the box floor.  The suspended rod hangs between the aluminum plate and the force sensors and is oriented with its long axis parallel to the plate long axis.

Two rows of cylinders are glued to a stationary steel block, as shown in Figure~\ref{fig:cohesionSchematic}, one for measuring the forces between pairs of cylinders and one for measuring the interactions between triplets of cylinders.  The setup for pairwise measurements corresponds to the stationary block oriented such that the single white star is in the upper right corner, as shown in Figure~\ref{fig:cohesionSchematic}.  For each interacting pair, the line connecting the centers of the cylinders is parallel to the direction of driving, indicated by $\vec{v}$ in Figure~\ref{fig:cohesionSchematic}.  An example of the geometry for a single cylinder pair viewed from the side is shown in Figure~\ref{fig:capRise}b.  For triplets, the stationary block is oriented such that the two white stars are in the upper right corner.  Each cylinder attached to the suspended rod interacts with two cylinders on the stationary block, forming equilateral triangles when the two rows are in contact.  An example of the geometry for a single cylinder triplet viewed from the side is depicted in Figure~\ref{fig:capRise}c.

Cylinders are placed into contact after oil has been added to the surrounding box.  Once capillary bridges have formed between the interacting sets of cylinders, the aluminum plate is then driven backward at $0.017$~mm/s.  The suspended rod resists this driving when capillary bridges are present and is therefore pushed against the force sensors, which are moving with the aluminum plate.  The  resulting force, measured as a function of displacement, $y$, is the sum of the individual capillary forces simultaneously acting on each of the cylinder pairs or triplets.  

\subsection{Numerical Calculations}
Numerical computations are performed using Surface Evolver~\cite{Brakke:1992tn,BrakkeBook}, a finite element modeling software package.  Once the configuration geometry is defined along with relevant physical parameters and constraints, Surface Evolver uses the method of gradient descent to iteratively evolve the fluid surface toward the minimum total energy state.  The fluid surface is represented by triangular elements, the size and density of which can be adjusted in between evolution steps. 

For each computation, the configuration of upright cylinder pairs or triplets is defined by specifying the cylinder separations and exposed heights above the fluid.   The undisturbed fluid resides in the $z=0$ plane, and the exposed cylinder height, $h$, is varied by adjusting the height of the cylinders above the $z = 0$ plane.  The size of the surrounding box containing fluid is set to be $20 R$ and is kept constant in all configurations.  Constraints on the fluid-cylinder surface prevent the fluid from penetrating cylinder walls and constraints at the edges of the box fix the fluid vertices to $z=0$.  An additional constraint is imposed at the fluid-cylinder surface to model the interactions between cylinders of finite height: the fluid vertices in contact with the cylinders are not allowed to exceed the exposed cylinder height.  

For a given exposed cylinder height, $h$ (see Figure~\ref{fig:capRise}a), configurations are defined with surface separations ranging from $d = 0.01$~mm to $d = 10.0$~mm, and a separate energy-minimization is performed for each configuration.  The treatment of each separation as an independent minimization is valid in the quasi-static limit of cylinder separation, which holds for slower separations speeds.  Within a given configuration, the cylinder positions are fixed and only the fluid is allowed to evolve.  The fluid is initially a flat surface in the $z = 0$ plane.  After a few mesh refinements, each of which divides each fluid element into four new elements, and a few evolution iterations, each of which moves the fluid surface to a lower-energy configuration, the fluid begins to rise up between the cylinders.  Triangle elements with area less than $5\times10^{-13}$~m$^2$ are regularly removed from the mesh to prevent instability of the gradient descent method.  Once the fluid motion becomes small, indicating the capillary rise has nearly reached the equilibrium rise height, then the surface is further refined and evolved until the energy difference between successive iterations,  $\Delta E $, is on the order of $10^{-13}$~J and the relative energy change between successive iterations is $\Delta E / E \sim 10^{-10} $.  Examples of the minimized surfaces are depicted in Figure~\ref{fig:capRise}b for pairs and  Figure~\ref{fig:capRise}c for triplets.  

\subsection{Theoretical Model}
\begin{figure}[h!]
 \centering
  \includegraphics[width=5in]{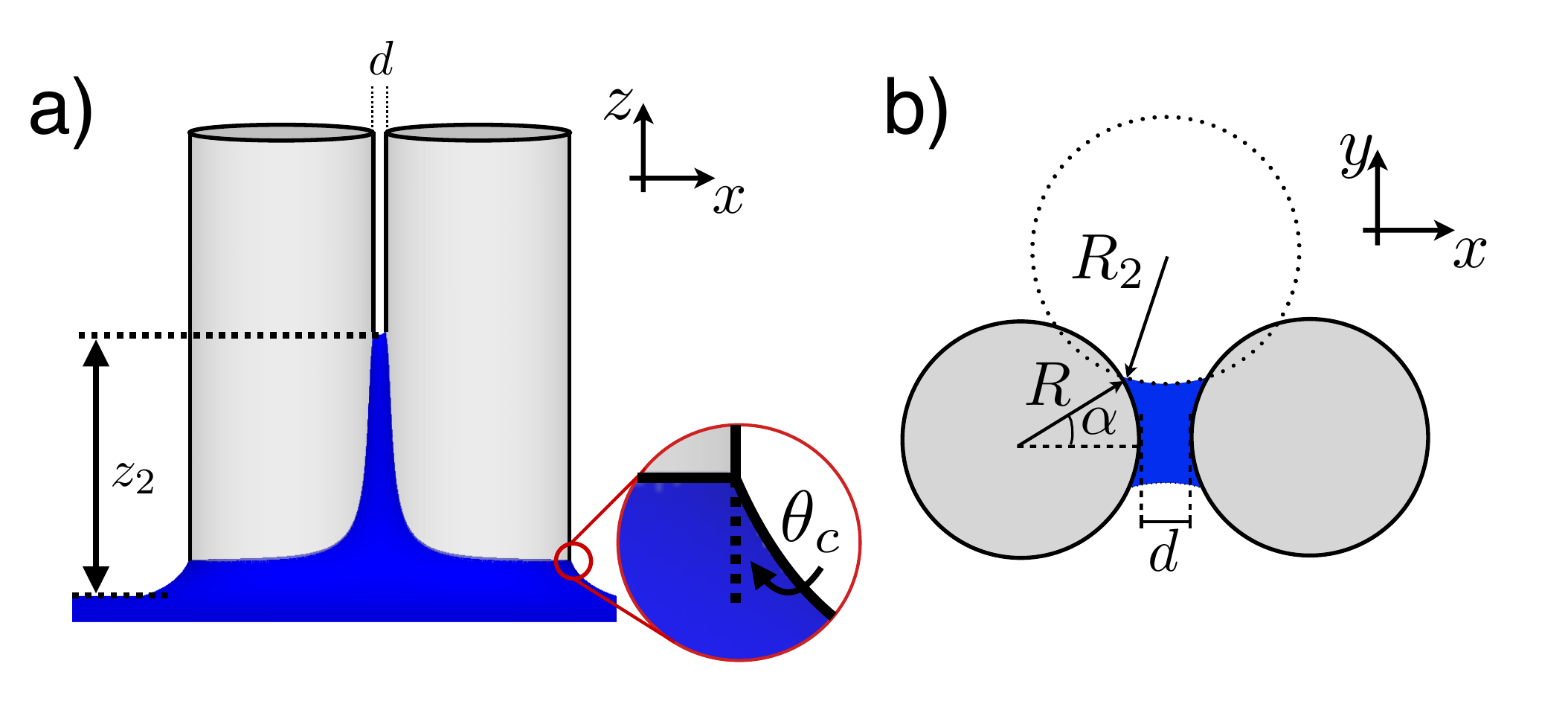} 
  \caption{a) Side view of a capillary bridge between two upright cylinders, where $z_2$ is the equilibrium rise height for a fluid with contact angle $\theta_c$ between two cylinders, each of radius $R$,  with surface separation $d$.  b)  View of the horizontal slice in the $z = z_2$ plane. $R_2$ is the radius of curvature of the fluid free surface in this horizontal plane, and $\alpha$ is the angle between the line connecting the centers of the cylinders and the line from the center of the cylinder to the contact line. }
  \label{fig:model}
\end{figure}

We aim to develop a model for the lateral capillary forces between upright cylinders of finite height, such as those shown in Figure~\ref{fig:model}, that will provide insight into the origin and relative importance of various contributions to the total attractive force.  To understand and characterize these lateral forces, we need to determine the region of the cylinder over which fluid forces are acting, and then integrate local lateral forces over this region to determine the total attractive force between two cylinders.  

Princen~\cite{Princen:1969ty} developed a model to estimate the capillary rise between two vertical cylinders of infinite height partially submerged in a fluid.  This model assumes that the capillary rise height, $z_2$, is much greater than the cylinder radius, $R$.  In this regime, changes in the vertical curvature of the fluid are small, so fluid between the cylinders is treated as a perfectly vertical column wherein the horizontal cross-section of the fluid at height $z$ is equal to the cross-section at $z_2$ for all $z$.  It follows from this assumption that the geometry of the system can be completely described by a horizontal cross-section of the fluid, as shown in Figure~\ref{fig:model}b, and that the fluid radius of curvature in the vertical direction is infinite.  This approach permits the hydrostatic pressure across the fluid interface to be described completely by the horizontal radius of curvature, $R_2$, 
\begin{align} 
\gamma / R_2(z,d) = \rho g z,  
\label{eqn:laplace}
\end{align}
where $\gamma$ is the surface tension, $\rho$ is the fluid density, $g$ is the acceleration due to gravity, and $z$ is the height from which the cross-section is taken.

We begin, as Princen did, by estimating the capillary rise height from vertical force balance.  The total vertical force must vanish at the equilibrium capillary rise height, so we solve the following equation for the capillary rise height, $z = z_2$, at which the weight of the fluid between the cylinders is equal to the surface tension forces acting at the interfaces: 
\begin{align} 
z \rho g A(z,d) = 4\gamma R \alpha(z,d) \cos[\theta_c] - 4 \gamma [\pi/2 -\theta_c - \alpha(z,d) ] R_2(z,d),
\label{eqn:fvert}
\end{align}
where $\theta_c$ is the contact angle, $A(z,d)$ is the area of a horizontal cross-section of the fluid, $d$ is the separation between cylinder surfaces, $z$ is the height from which the horizontal cross-section is taken, and $\alpha(z,d)$ is the the horizontal angle between the line connecting the cylinder centers and a line from the center of a cylinder to the contact line on the surface of that cylinder, see Figure~\ref{fig:model}b.  Expressions for $A(z,d)$ and $R_2(z,d)$ can be determined from geometry~\cite{Princen:1969ty}.
 
We use Eq.~(1) and the expression for $R_2(z,d)$ to determine $\alpha(z,d)$, which, when substituted into Eq.~(2) along with expressions for $A(z,d)$ and $R_2(z,d)$, yields a transcendental equation that can be numerically solved for the capillary rise height, $z_2$:
\begin{align} 
 0 = \Bigr(\frac{\gamma}{\rho g z_2 R}\Bigr)^2& \Bigr\{\frac{\pi}{2} - \theta_c + \alpha(z_2,d) + \sin[\theta_c + \alpha(z_2,d)] \cos[\theta_c + \alpha(z_2,d)]\Bigr\}  \nonumber \\
& +2 \frac{\gamma}{\rho g z_2 R} \Bigr\{\sin[ \alpha(z_2,d)] \cos[\theta_c + \alpha(z_2,d)] - \alpha(z_2,d) \cos[\theta_c] \Bigr\} \nonumber \\
& +\sin[\alpha(z_2,d)] \cos[\alpha(z_2,d_2)] -\alpha(z_2,d).
\label{eqn:trans}
\end{align} 
Unlike Princen, our cylinders have finite height, $h$, above the liquid in which they are partially submerged.  If the calculated $z_2$ exceeds $h$, as it often tends to for small separations, we set $z_2 = h$.  

Using this capillary rise height, we can estimate the lateral attractive force between the cylinders. Both pressure and surface tension contribute to the total capillary-induced attractive force on one cylinder, which is given by
\begin{align} 
F_{\mathrm{total}}(z_2,d)  = 2 & \int_0^{z_2}  \rho g z R \sin[\alpha(z,d)] dz \nonumber \\
&+ 2 \gamma R \sin[\theta_c] \sin[\alpha(z_2,d)] + 2 \gamma  \int_0^{z_2} \sin[\alpha(z,d) + \theta_c] dz.
\label{eqn:ftotal}
\end{align} 
The first term is the pressure contribution, which acts over the cylinder-fluid contact area, the second term comes from the surface tension acting at the fluid-air-cylinder interface along the top of the capillary bridge, and the third term arises from the surface tension acting at the fluid-air-cylinder interface along the height of the cylinder. The model allows us to examine the relative importance of the independent force contributions, something we do not have access to from our other measurements or calculations.

Note we relax the $z_2 \gg R$ assumption only after determining the rise height, $z_2$.    We allow the horizontal cross-section to vary with $z$ for the lateral force calculation, while at the same time still assuming that the each horizontal cross-section can be treated independently and summed over to yield the total attractive force.  Despite this technical inconsistency, we show in the results section that the total force from this model agrees well with experimentally-measured and numerically-computed forces.  

\section{Results and Discussion}
\subsection{Interactions Between Cylinder Pairs}
\begin{figure}[h!] 
 \centering
  \includegraphics[width=5in]{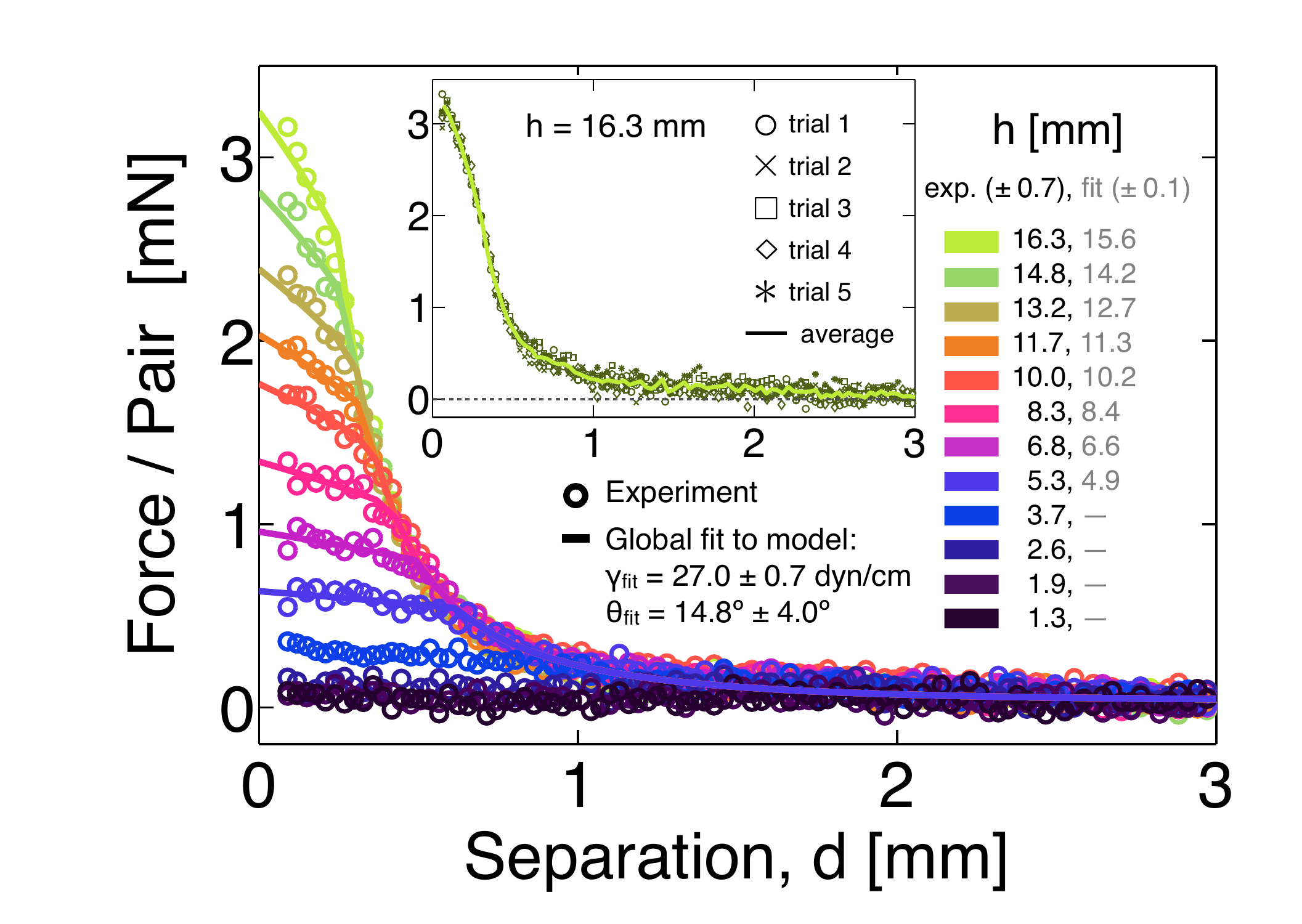} 
  \caption{Experimentally measured attractive forces between cylinder pairs (circles) as a function of separation for many exposed cylinder heights, $h$ as labeled.  Each curve is the result of averaging five experimental trials together.  Raw data and the resulting average curve for $ h = 16.3$~mm are shown in the inset. The uncertainty in the depth measurements is $\pm0.7$~mm, and forces for $d < 80$~$\mu$m are excluded because they cannot be measured reliably. Lines represent the result of a global fit of the model to the largest eight largest exposed cylinder heights.  }
  \label{fig:pairsExp}
\end{figure}

Averaged force curves for many different exposed cylinder heights, $h$, can be seen in the main plot of Figure~\ref{fig:pairsExp}.  Due to the nature of the setup, we cannot reliably measure forces for separations smaller than $80$ $\mu$m.  While the cylinders are initially in contact, it takes a finite but small amount of time for the suspended rod to establish full contact with and subsequently push on the force sensors.  Once this occurs, the forces quickly jump to a maximal value and then slowly decrease as separation increases.  As a result, we exclude force data for separations smaller than $d = 80$ $\mu$m, which is the separation at which this maximal force occurs.      

The capillary bridges reach the tops of the cylinders for small surface separations, causing the force to deviate from the infinite-height cylinder predictions at small separations.  This effect causes the force to depend on the height of the exposed cylinder above the oil.  Not surprisingly, the maximum attractive force is greater for larger exposed cylinder heights.  Forces for all $h$ collapse at large separations, which is expected.  The dependence of the forces on $h$ arises when the equilibrium capillary rise height exceeds the cylinder height.  At large separations capillary rise height never reaches the cylinder tops, causing the dependence on $h$ to vanish.  In the intermediate-separation regime, forces for larger cylinder height $h$ collapse at lower $d$ than corresponding forces for smaller $h$.  This effect is also expected, because the capillary bridge height will fall below $h$ sooner for larger exposed cylinder heights.

We perform a global fit of the model discussed in the Methods section to the data for the eight largest $h$ values and extract the surface tension and contact angle that best describe the data, as well as the $h$ value that best fits each of the eight data sets.  The resulting best-fit contact angle is $\theta_{\mathrm{fit}} = 14.8^{\circ} \pm 4.0^{\circ}$ and the surface tension is $\gamma_{\mathrm{fit}} = 27.0 \pm 0.7$~dyn/cm, both of which are close to the experimentally measured values.  The values for each of the exposed cylinder heights, $h_{\mathrm{fit}}$, are within the uncertainty of experimentally measured values and have $95\%$ confidence intervals of $\pm 0.1$~mm.  Given the smaller bounds on the fit values for the exposed cylinder heights, we use the model fit parameters in all numerical computations.

\begin{figure}[h!] 
\centering
  \includegraphics[width=5in]{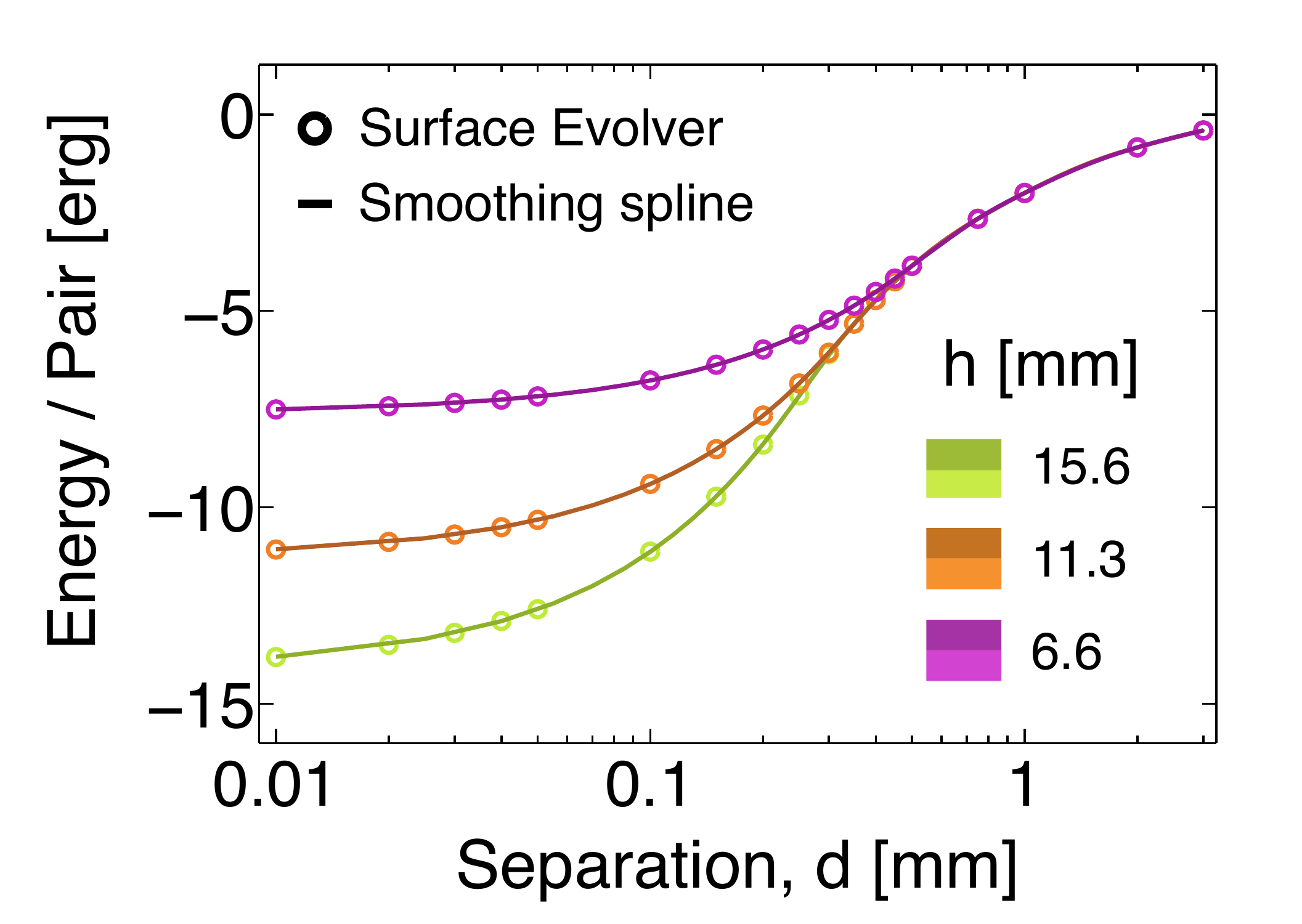} 
  \caption{Surface energies as a function of cylinder separation determined using Surface Evolver, using best-fit parameters from the model.  Each data point (circle) corresponds to one simulation.  The simulation data for each exposed cylinder height is fit to a smoothing spline (solid lines), which is then differentiated to determine the force of attraction between cylinder pairs as a function of $d$, the separation of the cylinder surfaces.    }
  \label{fig:pairsSE}
\end{figure}

Fluid-mediated interactions between a pair of upright cylinders are also explored numerically using Surface Evolver~\cite{Brakke:1992tn,BrakkeBook}.  The minimized energy values as a function of separation are shown in Figure~\ref{fig:pairsSE} for three values of exposed cylinder height.  Each open circle is the result of an energy-minimization calculation for a given surface separation, $d$, and exposed cylinder height, $h$.  Smoothing splines are fit to each data set for a particular $h$ and differentiated to obtain the attractive forces between the cylinders.

\begin{figure}[h!]
  \centering
  \includegraphics[width=5in]{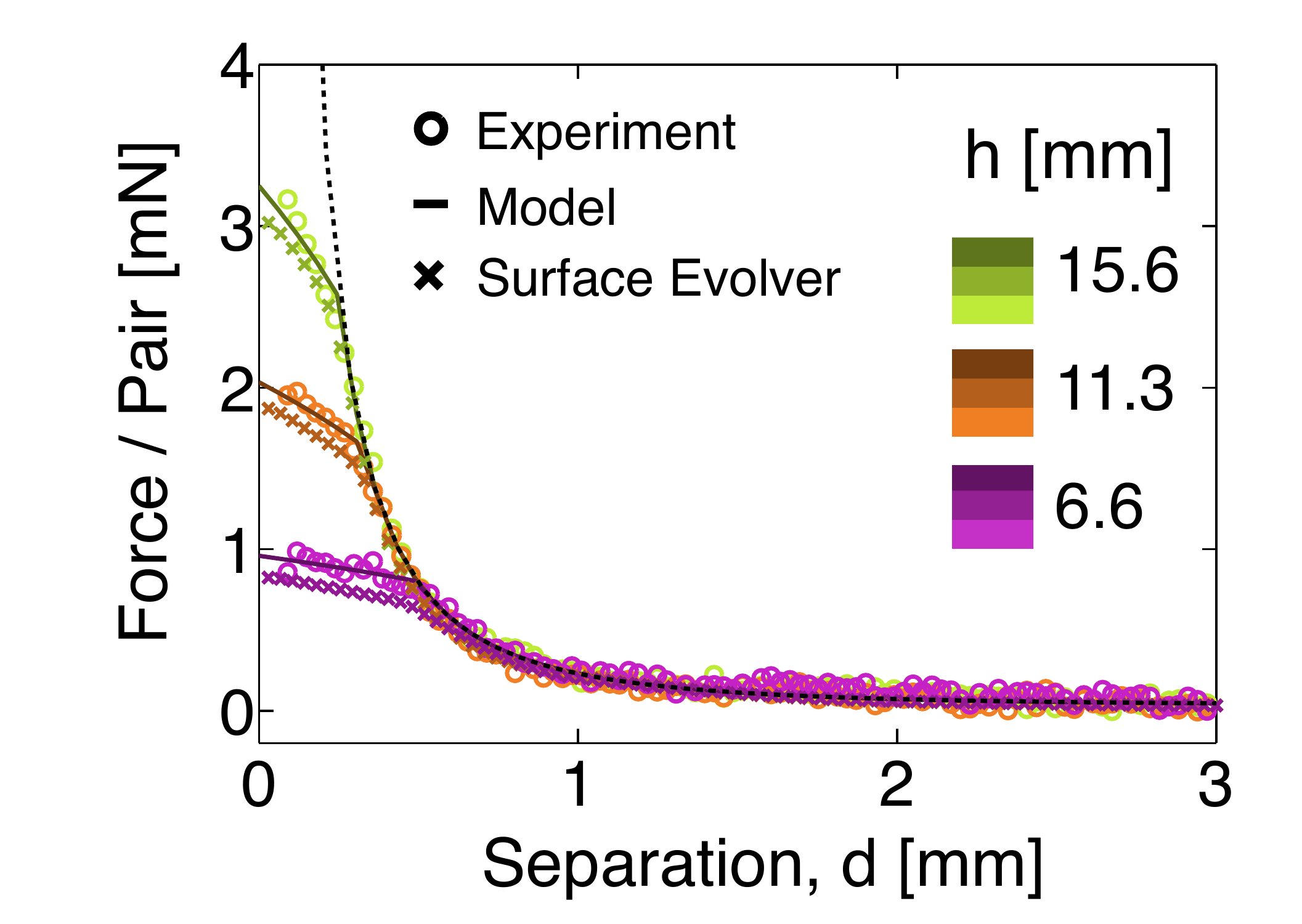} 
  \caption{Attractive forces between cylinder pairs: experimental measurements (circles), Surface Evolver energy derivatives (lines), and model calculations (squares) for three different liquid levels.  The dashed line indicates model-predicted force for infinitely tall cylinders.  Deviations from this line occur at small separations, $d$, because the capillary rise has reached the tops of the cylinders.   }
  \label{fig:pairsAll}
\end{figure}

The theoretical model discussed in the previous section divides the total attractive force into two contributions: a force due to hydrostatic pressure inside the fluid and a force due to surface tension along the air-fluid-cylinder interfaces.  The resulting attractive force predictions are compared for three exposed cylinder heights in Figure~\ref{fig:pairsAll}.  The corresponding experimental measurements and numerical computations are shown as well, and all are in reasonable agreement.  The dashed line on the plot shows the result of the model for infinitely tall cylinders.  Deviations from this line at small separations, $d$, are caused by the finite cylinder height, specifically when the equilibrium capillary rise height exceeds the cylinder height.  While the cylinder height does not explicitly enter into the model calculation, it is imposed by not allowing the capillary rise height to exceed the cylinder height.  The resulting maximum fluid heights are consistent with those measured from the final states of Surface Evolver calculations, and the resulting force curves capture reasonably well the small-separation behavior observed in experimental measurements as well as numerical calculations.  

\begin{figure}[h!] 
  \centering
  \includegraphics[width=5in]{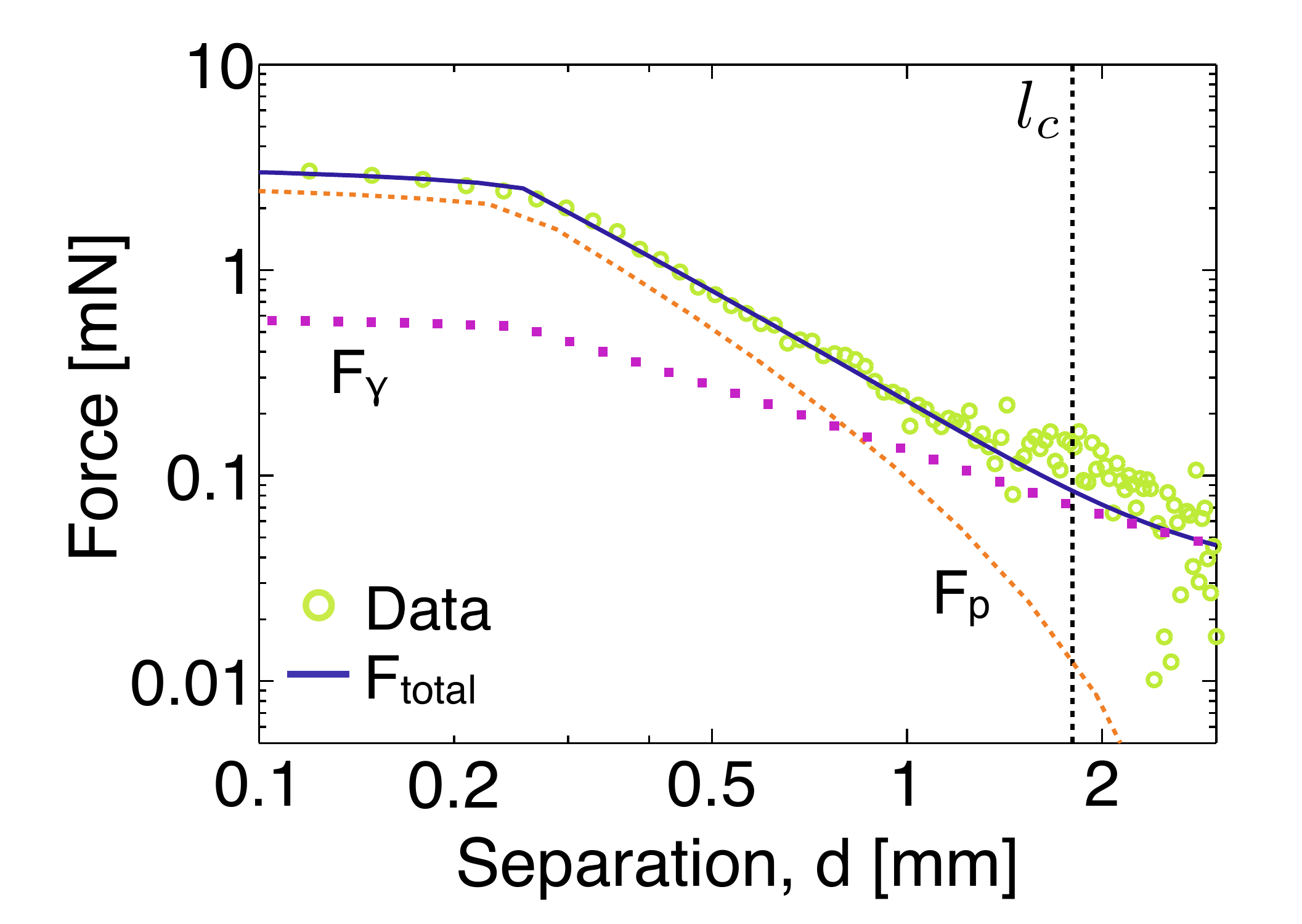} 
  \caption{Experimental data for $h=16.3$~mm shown (circles) along with total predicted force from the model using best-fit parameters from the model (solid line).  The total predicted force, given in Eq.~(4), is the sum of three terms.  The first term in Eq.~(4) is the force contribution from the pressure inside the fluid (dashed line) and remaining two terms describe the surface tension force (dotted line).  The force is dominated by pressure for small separations, while surface tension is more important for large separations.  There is a cross-over in the dominant contribution to the total force around $0.5 l_c$. The near-plateau at small separations, $d$, is due to the finite height of the cylinders.  In this entire region, the capillary rise height reaches the tops of the cylinders, so the increase of force here is caused only by an increase in the thickness of the capillary bridge as the separation between the cylinders decreases.}
  \label{fig:fcomp}
\end{figure}

The model provides insight into the relative importance of the surface tension and pressure force terms.  Figure~\ref{fig:fcomp} shows experimentally measured forces for one exposed cylinder height, along with both the total force predicted by the model as well as the individual components that contribute to the total force. The pressure term dominates the force at small separations, there is a crossover around $d = 0.5 l_c$, and then surface tension dominates for $d \gtrsim l_c$.  One limitation of the model is that breaks down for $d > 2 l_c$; thus we are unable to predict how these forces behave at very large separations.

\subsection{Interactions Between Cylinder Triplets}
\begin{figure}[h!]
\centering
  \includegraphics[width=5in]{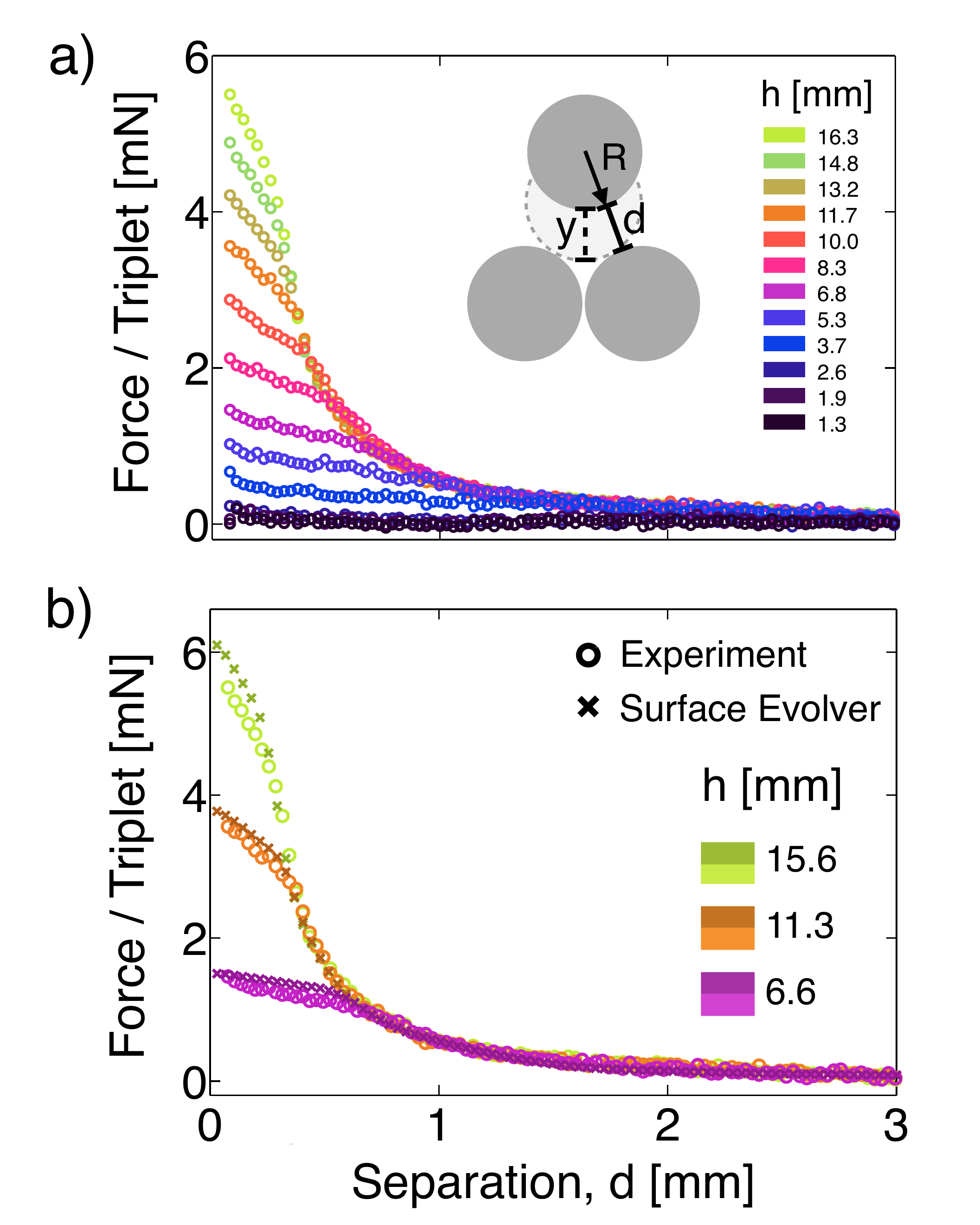} 
  \caption{a) Force vs separation for a group of three cylinders for many exposed cylinder heights, $h$.  b) Experimental force data and Surface Evolver energy derivatives (using parameters from model fit) show reasonable agreement for three different $h$ values. }
  \label{fig:triplets}
\end{figure}

To test pairwise additivity, we also measure the force required to pull one cylinder away from two neighbors, with all three initially in mutual contact.  The setup is similar to the one depicted in Figure~\ref{fig:cohesionSchematic}, the only difference being that the stationary block is rotated $180^{\circ}$ so that the two white stars are in the upper right corner.  Forces are measured as a function of aluminum-plate displacement, $y$, using the same procedure as for pairs.  For triplets, however, the surface separation, $d$, is not equivalent to $y$, though they are geometrically related through the equation $d = -2R+\sqrt{4R^2 + 2\sqrt{3} R y + y^2}$.  Figure~\ref{fig:triplets}a shows the final force curves, each of which is the average of five independent experiments, as a function of $d$ for numerous exposed cylinder heights.  

Surface Evolver is used to numerically determine the minimum energy of a fluid surface disturbed by the presence of three upright cylinders.  Energy minimizations are performed for numerous configurations, such as the one shown in Figure~\ref{fig:capRise}c, each with fixed values of  $h$ and $y$.  For each value of $h$, a smoothing spline is fit to corresponding energy data points and differentiated to obtain the attractive forces between cylinder triplets. Figure~\ref{fig:triplets}b shows reasonable agreement between differentiated Surface Evolver energies for three values of $h$ and the corresponding experimental data.  

We can determine the importance of non-pairwise terms to the overall force by comparing the pairwise and triplet force data.  The forces measured for each cylinder triplet have a contribution from the capillary bridges between two cylinder pairs as well as the liquid that rises up in the center of the triangle formed by the three cylinders.  An example of these capillary bridges can be seen in the final state of a Surface Evolver energy minimization in Figure~\ref{fig:capRise}c as though viewed from the side, through translucent cylinders.  

\begin{figure}[h!]
  \centering
  \includegraphics[width=5in]{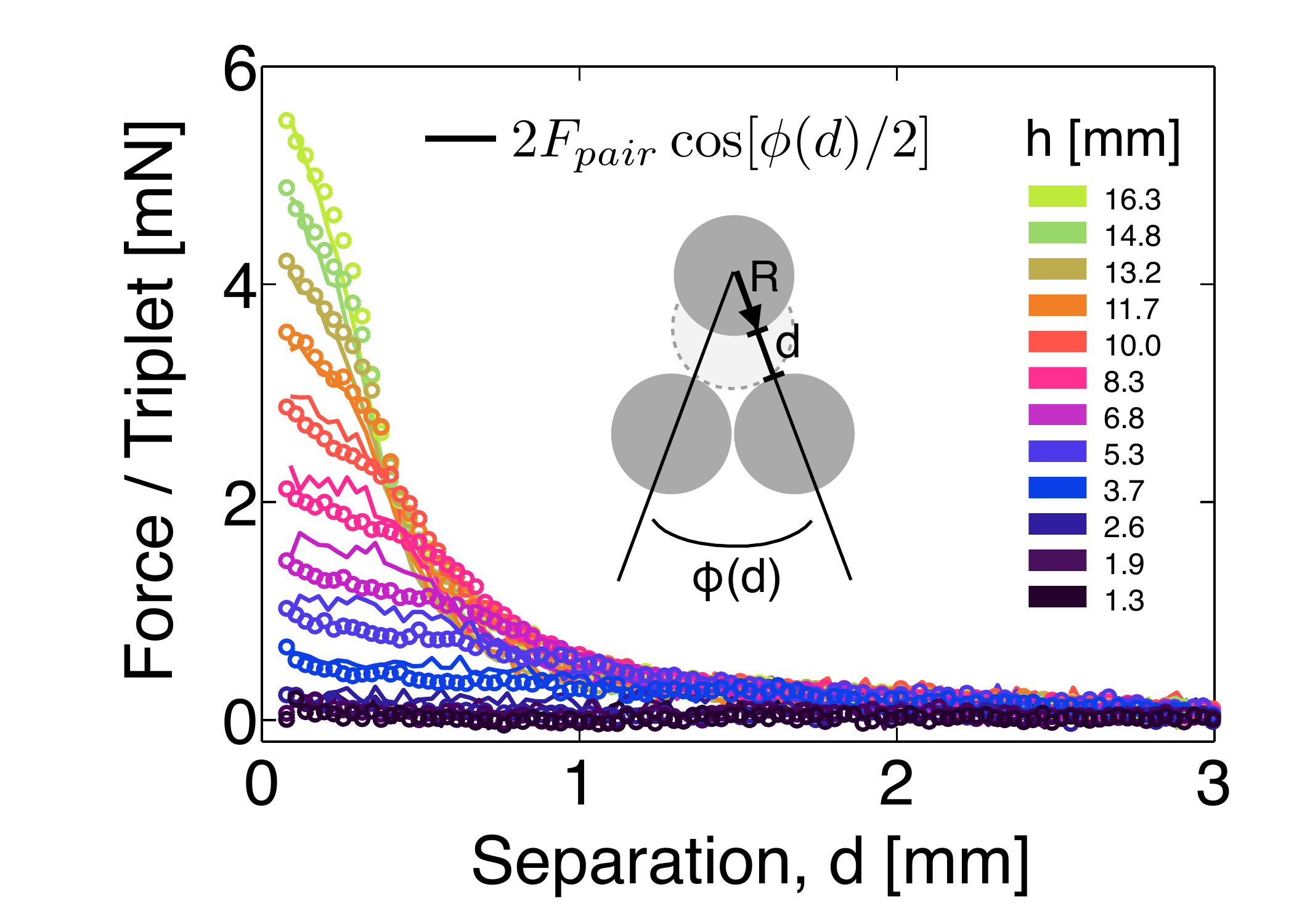} 
  \caption{Force vs separation data are shown for triplets (circles) for numerous exposed cylinder heights.    Forces between triplets will have contributions from the two pair-wise interactions acting in the directions $ \pm \phi(d)/2 = \pm \arctan(R/\sqrt{d^2 + 4Rd +3R^2})$ relative to the direction of separation as well as a contribution from a capillary rise that occurs in the middle of the three cylinders.  Contributions expected from the two pairwise interactions (solid lines)  account for nearly all of the measured triplet interactions.   }
  \label{fig:pairsAndTriplets}
\end{figure}

We compare triplet forces with the expected forces for two interacting pairs in Figure~\ref{fig:pairsAndTriplets}.  To make this comparison, we must account for the fact that the force sensors are only measuring the component of the force in the direction of the motion.  For the measurements between cylinder pairs, the direction of the maximum force and the direction of motion are the same.  For the triplets, however, these directions differ by the angle $\phi(d)/2 = \arctan(R/\sqrt{d^2 + 4Rd +3R^2})$, so we compare $F_{\mathrm{triplet}}$ to $2 F_{\mathrm{pair}} \cos[\phi(d)/2]$ in Figure~\ref{fig:pairsAndTriplets}.  Forces between triplets are reasonably well-described by the pairwise interactions, though the pairwise data falls off a bit faster in the $0.5$ to $1$~mm range. Discrepancies for low $h$ may be due to the $\pm 0.7$~mm uncertainty in the depth measurements.  The overall agreement indicates that the contribution from lower capillary bridge in the center of the three cylinders is comparatively small and can be neglected.  The overall agreement indicates that the contribution from the lower capillary bridge in the center of the three cylinders is comparatively small and can be neglected.  We expect that the capillary rise will be even smaller in the center of four or more cylinders and that, therefore, pairwise additivity is a reasonable approximation for arbitrary configurations of upright cylinders.

\subsection{Hysteresis Between Cylinder Pairs}
Contact angle dependence on the velocity of the contact line has long been observed in systems with relative motion between a solid and a fluid~\cite{Ablett:1923cz,Bartell:1940ua,YARNOLD:1946ud,ROSE:1962tm,Elliott:1962tf, Elliott:1967wi,Hoffman:1975ub,JohnsonJr:1977ti,Dussan:1979uk,JohnsonJr:2009esa,STROM:1990tx, Shikhmurzaev:1996wj,Ranabothu:2005fa,Tavana:2006ks,Keller:2007iw}.
The advancing contact angle,  $\theta_A$, measured when the fluid-solid contact area increases, is always measured to be greater than the receding contact angle,  $\theta_{R}$, which is measured when the fluid-solid contact area decreases.  $\theta_A$ is observed to increase with increasing speed, and $\theta_R$ has been observed to decease with increasing speed in some experiments, though the $\theta_R$ data tends to be more scattered.
\begin{figure}[h!]
  \centering
  \includegraphics[width=5in]{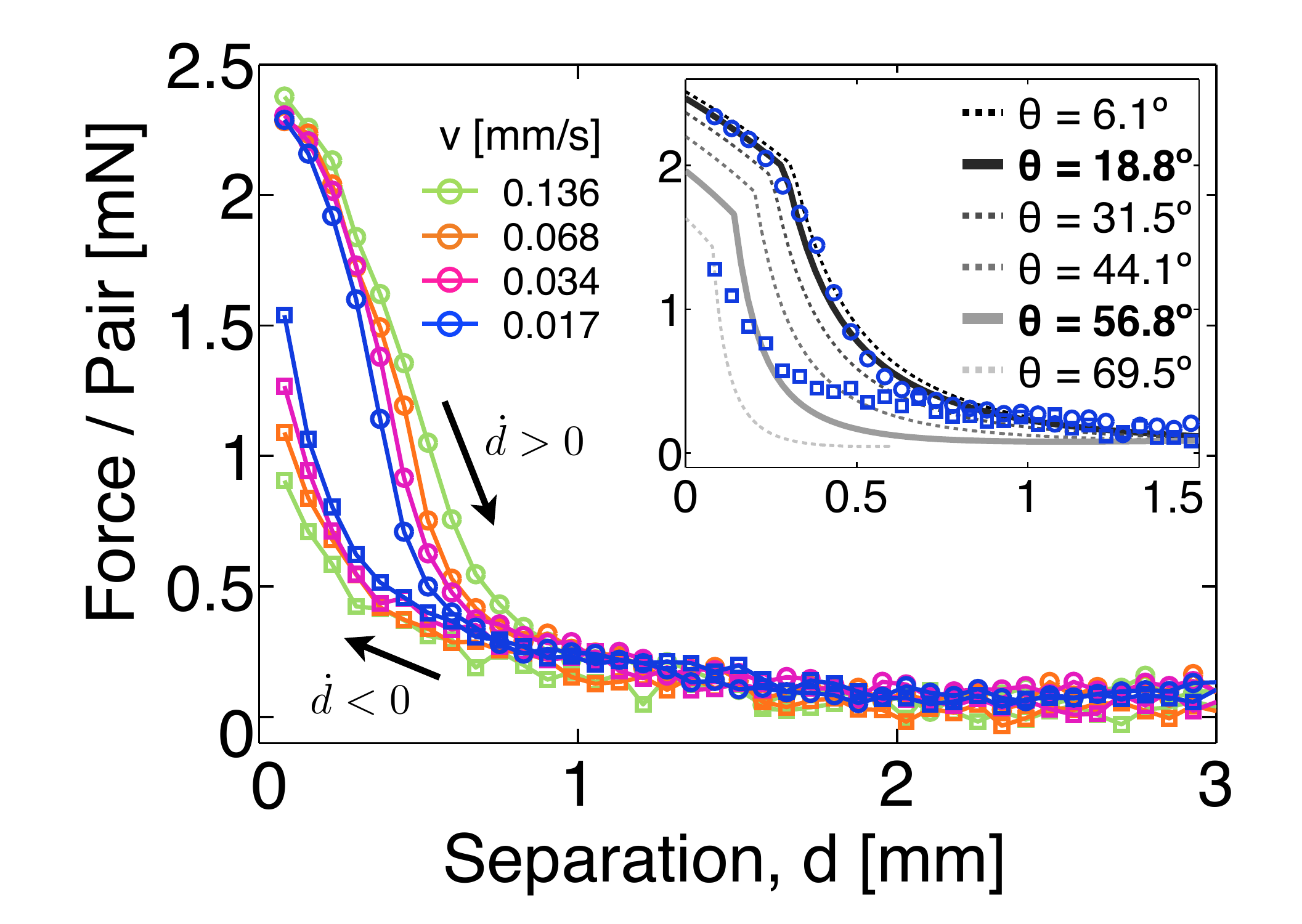} 
  \caption{Measured force vs separation for cylinder pairs at different speeds for $h = 14.1\pm 0.7$~mm.  A direction-of-motion-dependent hysteresis is observed, the strength of which is dependent upon the speed of the motion.  The top curves (circles) are measured when the cylinder separation is increasing and the bottom curves (squares) are measured as the cylinders are pushed together.  Lines represent the average of five to fifteen experiments, and the size of the points is indicative of the uncertainty. In the inset, experimental data for the slowest speed is plotted along with six evaluations of the model.  The best-fit force curves are shown as the solid lines.  Both contact angle and $h$ are fit parameters for increasing separation data (circles).  For the decreasing separation data, $h$ is fixed and contact angle is the only fit parameter.  }
  \label{fig:hysteresis}
\end{figure}

The experimental setup used to measure attractive forces between cylinder pairs, shown in Figure~\ref{fig:cohesionSchematic}, is also used to characterize the hysteresis in these attractive forces.  Cylinders are initially placed into contact and, after capillary bridges form between all fifteen cylinder pairs, the aluminum plate is driven away from the stationary block to a distance of $6$~mm.  The aluminum plate remains static for one minute, after which the plate is driven back to its original position at the same speed. Averages of at least five experiments for each of four different speeds are shown in Figure~\ref{fig:hysteresis}.  For small separations, forces measured while increasing separation are always larger than the corresponding forces measured for decreasing separation. Forces at larger separations do not depend on the direction of driving, indicating that drag forces on the cylinders, and lubrication forces between the cylinders and the base of the surrounding box, are not responsible for the observed hysteresis.

Our data is qualitatively consistent with previous work\cite{Velev:1993wz}, in which the  forces between approaching sub-millimeter cylinders at a separation of $0.5 l_c$ were found to be $10-15\%$ smaller than the corresponding forces between separating cylinders.  We also observe forces measured during separation to be higher than those measured while pushing cylinders together, though the magnitude of this difference is speed-dependent.  This hysteresis in the measured forces is also qualitatively consistent with what is known about contact angle hysteresis.  The contact line is receding down the cylinder surface when the separation between cylinders is increasing.   This reduces the contact angle and leads to an increased force.  Similarly, the contact line is advancing up the cylinders when cylinder separation is decreasing, causing an increase in the contact angle and leading to a decrease in the measured force.  

The hysteresis measurements for the slowest speed are compared to the model in the inset of Figure~\ref{fig:hysteresis}.  In these experiments, the exposed cylinder height is measured to be $h = 14.1\pm 0.7$~mm, and the static contact angle is estimated to be  $\theta = 20^{\circ} \pm 5^{\circ}$ from numerous photographs of a single cylinder in oil. The surface tension, measured to be $\gamma = 27.6 \pm 0.9$~dyn/cm, is determined from photographic measurements of capillary rise heights inside capillary tubes of both $5$ $\mu$L and $50$ $\mu$L volumes.   Using these experimentally measured values, we simultaneously fit the model to increasing separation data at speed $v = 0.017$~mm/s for two different liquid depths. Given the uncertainty in $h$, we allow both $h$ and $\theta$ to vary in the fitting, and find the best-fit contact angle to be $\theta_{R\mathrm{,fit}} = 18.8^{\circ} \pm 2.0^{\circ}$ and the best-fit exposed cylinder height for the data shown in the inset of Figure~\ref{fig:hysteresis} to be $h_{\mathrm{fit}} = 12.9 \pm 0.3$~mm.   We then fix the fit parameter $h_{\mathrm{fit}}$ to find the best-fit contact angle for the decreasing separation cylinder data, $\theta_{A\mathrm{,fit}} = 56.8^{\circ} \pm 2.7^{\circ}$. 

The best-fit model force curves are shown as the thick solid lines in the inset Figure~\ref{fig:hysteresis}, and curves from two intermediate angles, as well as one below $\theta_{R\mathrm{,fit}}$ and one above $\theta_{A\mathrm{,fit}}$, are shown as dashed lines to give a sense of the model force dependence on the contact angle.  The dark solid line is the result for the increasing separation data, and the fit captures the behavior of the experimental data pretty well.   The best fit of the model to the decreasing separation data, shown as the light solid line,  does not describe the data well, which perhaps indicates that the decreasing cylinder separation forces cannot be described by a single contact angle.

The speed dependence of the separating cylinder data can be seen in Figure~\ref{fig:hysteresis}.  As the speed increases, the force curves become broader and the forces fall off more slowly, especially for the two fastest speeds.  Comparing this data with the model behavior in the inset, we see that a smaller contact angle is not enough to account for the changes observed in the force curves, indicating perhaps that the quasi-static assumption is not valid at faster speeds.

\section{Conclusion}
In this paper, we have characterized capillary-induced attractive forces between millimeter-sized cylinder pairs and triplets.  Experimental measurements made with a custom-built apparatus are in reasonable agreement with numerical computations and a simple theoretical model.  The model enables us to ascertain the surface tension and pressure contributions to the total force separately, and therefore compare their relative importance.  We find that, at small separations, the pressure term dominates the total force, and at large separations, the surface tension force dominates. 

The forces between triplets are reasonably well-described by the pairwise interactions.  While some small discrepancies between the triplet and scaled-pair forces were found, we expect that these will monotonically decrease as the number of cylinders is increased.  Therefore, pairwise additivity is a reasonable approximation for descriptions of the forces in a system with similar physical parameters and an arbitrary number of cylinders.  

We also observed a velocity-dependent hysteresis of force measurements between cylinder pairs.  For separations less than $1$ mm, forces measured while separating cylinders are always larger than the corresponding forces measured for approaching cylinders. This finding is qualitatively consistent with previous observations.  The size of the hysteresis is observed to increase with increasing speed.  We demonstrate that the simple model does not fit the data when the cylinder surfaces are approaching one another, which may suggest that a single contact angle is not enough to describe the data.  We also show that the speed dependence of the separating cylinder data is not described by the model, perhaps indicating that the quasi-static assumption is no longer valid for the faster speeds.    

Lastly, we observe that when the capillary rise height is greater than the cylinder height, the attractive force between cylinders depends the height of the cylinder above the liquid level.  This effect can be employed to create a tunable cohesion.  One benefit of such a force is that the liquid is distributed evenly throughout an array of cylinders or other particles, so that the force of attraction is known everywhere.  

\begin{acknowledgement}

We thank Dominic Vella and Cesare M. Cejas for helpful discussions.   This work was supported by the National Science Foundation through Grant MRSEC/DMR-1120901.

\end{acknowledgement}

%
%

\bibliography{cohesion_langmuir_arXiv.bib}

\end{document}